# Advances in Biomedical Devices: A comprehensive Exploration of Cardiovascular and Ophthalmic Applications

Shankar M
Healthcare Technology Innovation
Indian Institute of Technology Madras
Chennai, India
shankar.m@htic.iitm.ac.in

*Abstract*—This review article discusses current technological advances in biomedical devices, emphasizing cardiovascular and ophthalmic applications—diagnostic, monitoring, and prosthetic instruments and systems. The scope encompasses various aspects, including implantable retinal prosthetic devices, portable devices for carotid stiffness measurement, automatic identification algorithms for arteries, cuffless evaluation of carotid pulse pressure, wearable neural recording systems, and arterial compliance probes. Additionally, the paper explores advancements in pulse wave velocity measurement, real-time heart rate estimation from wrist-type signals, and the clinical significance of non-invasive pulse wave velocity measurement in assessing arterial stiffness. The synthesis of these studies provides insights into the evolving landscape of biomedical devices, their validation, reproducibility, and potential clinical implications, emphasizing their role in enhancing diagnostics and therapeutic interventions in cardiovascular and ophthalmic domains.

## I. Introduction

The rapid evolution of biomedical technologies has paved the way for ground breaking advancements in the realm of non-invasive devices, particularly in the cardiovascular and ophthalmic domains. This review explores the current landscape of technological innovations, shedding light on diagnostic, monitoring, and prosthetic instruments designed to enhance healthcare interventions in cardiovascular and ophthalmic applications. The discussed devices encompass a wide range of functionalities, including implantable retinal prosthetics, portable instruments for carotid stiffness measurement, automatic identification algorithms for arteries, cuffless evaluation of carotid pulse pressure, wearable neural recording systems, and arterial compliance probes. In the subsequent discussion, the review critically examines the validity and reproducibility of these non-invasive devices. Use the enter key to start a new paragraph. The appropriate spacing and indent are automatically applied.

Some key challenges and areas for further research include:
- Comparative Analysis: Evaluating methodologies in comparison to other established methods for comprehensive understanding of stiffness measurement.
- Cofounding factors: Considering and addressing the potential impact of confounding factors on measurement methodologies.
- Generalizability: Recruiting participants according to appropriate recommended guidelines to establish the robustness of the tested devices and methodologies.
- Clinical Practise: Validating and deploying new methodological devices into clinical practice on a large cohort for early diagnosis and lifestyle modification.

The review emphasizes the current state of these non-invasive biomedical devices in reshaping diagnostic and monitoring practices, contributing to early detection, continuous monitoring, and effective management of diseases.

## II. Literature Review

### A. Regional arterial stiffness

The pulse wave velocity (PWV) is estimated by measuring the time delay $\Delta t$ at the same point in the waveform (e.g., the foot of the wave) and estimating the arterial distance $\Delta D$ between the two measurement locations, using the formula $PWV = \Delta D/\Delta t$. The gold standard measurement of arterial stiffness is the carotid femoral PWV. Numerous user-friendly, portable biomedical devices have been developed for quick, non-invasive measurements of arterial stiffness, pulse wave velocity (PWV), and pulse pressure, utilizing modalities such as ultrasound and applanation tonometry. The importance of clinical validation, repeatability, and reproducibility persists, given the existence of numerous prototypes, as they are essential for establishing different reference values of PWV across diverse ethnic populations.

*1) Brachial-Ankle Pulse Wave Velocity*: A study was conducted to evaluate the alteration of baPWV in patients with coronary artery disease (CAD). Also, validity, reproducibility and clinical significance of non-invasive brachial-ankle pulse wave velocity (baPWV) measurements was discussed. The validity of baPWV was confirmed by comparing it with aortic pulse wave velocity (PWV) obtained using a catheter tip with pressure manometer, showing a strong correlation. The interobserver and intraobserver reproducibility of baPWV measurements were found to be considerably high, with correlation coefficients of 0.98 and 0.87, respectively, and coefficients of variation of 8.4% and 10.0%. The study also compared baPWV among patients with CAD, patients without CAD but with hypertension, diabetes mellitus, or dyslipidemia, and healthy subjects without these risk factors. The results showed significantly higher baPWV in CAD patients compared to non-CAD patients with risk factors, and higher baPWV in non-CAD patients with risk factors compared to healthy subjects without risk factors. The article concludes that the validity and reproducibility of baPWV measurements are high, and baPWV seems to be an acceptable marker reflecting vascular damages. The non-invasive baPWV measurement is suitable for screening vascular damages in a large population and has the potential to be used as a marker for cardiovascular risk assessment [1].

*2) Carotid-Femoral Pulse Wave Velocity:* The reproducibility of pulse wave velocity (PWV) and augmentation index (AIx) measured by pulse wave analysis (PWA) for use in large-scale clinical trials using SphygmoCor was investigated in this study. It involved two separate experiments to assess the reproducibility of PWV and AIx, where the wave transit time is calculated by the software, referencing the R wave of a simultaneously recorded ECG. The surface distance between the two recording sites is then measured, enabling the determination of PWV. It resulted that PWA is a simple and reproducible method for assessing arterial stiffness, with high reproducibility for AIx and lower reproducibility for PWV. It also highlights the potential importance of arterial stiffness as an independent measure of cardiovascular risk. Ultimately, the study suggests that PWA could improve risk stratification and provide valuable information about cardiovascular disease and its surrogate markers [2].

ARTSENS® Pen, a portable device with single-element ultrasound transducer for measuring carotid artery stiffness validation and clinical-utility assessment was investigated in another study. The device's performance underwent clinical validation on 523 subjects, utilizing a clinical-grade B-mode ultrasound imaging system (ALOKA eTracking) as the reference. Carotid stiffness measurements were conducted in a sitting posture, using the device. Key findings reveal a statistically significant correlation ($r > 0.80$, $p < 0.0001$) with a non-significant bias between the measurements obtained from the two devices. The results showed a strong correlation between measurements obtained from ARTSENS® Pen and the reference system, with non-significant bias. The device demonstrated high repeatability of measurements and sensitivity to detect age-related changes in arterial stiffness. The device's portability and automation make it suitable for field-level vascular screening. The study highlights the advantages of ARTSENS® Pen over traditional imaging systems in terms of usability, minimal expertise requirement, and ease of measurement procedures. Overall, the study provides comprehensive evidence to support the accuracy, repeatability, and clinical utility of ARTSENS® Pen for non-invasive measurement of arterial stiffness, making it a promising tool for cardiovascular risk assessment in various settings [3].

Blood pressure affects the values of PWV and the impact of blood pressure perturbations on arterial stiffness, focusing on the association between acute changes in blood pressure and arterial stiffness. The study involved 50 healthy subjects, including young and older adults, and utilized various blood pressure perturbation manoeuvres. The results revealed that all measures of arterial stiffness were influenced by acute changes in blood pressure, with varying degrees of dependency on blood pressure observed across different measures of arterial stiffness. The study also highlighted the potential influence of age and sex on the associations between changes in arterial stiffness and blood pressure, with stronger associations observed in men and younger individuals. Overall, the findings emphasize the significant impact of acute blood pressure changes on arterial stiffness and the variability in these associations across different measures of arterial stiffness and demographic groups [4].

*3) Pulse wave Imaging:* The feasibility of using Pulse Wave Imaging (PWI) to measure regional arterial stiffness in the human carotid artery was also investigated. The PWI technique involves using a high-frequency ultrasound to visualize the propagation of the pulse wave in the artery and estimate the regional PWV. The study reduced the beam density in the ultrasound image to increase the frame rate, allowing for better visualization of the pulse wave propagation. The regional PWV was estimated by tracking the wall velocities and analysing the spatiotemporal variation of the pulse wave. The results showed that PWI was feasible in the human carotid artery, with the estimated PWV values (4.0 to 5.2 m/s) being consistent with those reported in the literature. The study demonstrated the uniform propagation of the pulse wave along the carotid artery and highlighted the potential of PWI offering potential implications for the early detection and characterization of vascular diseases [5].

*B. Local arterial stiffness*

The devices mentioned earlier, requiring frequent calibration and uncomfortable pressure cuffs, are some of the challenges for regional Pulse Wave Velocity (PWV) measurement. Superficial arteries like the carotid site are considered easier for PWV measurement, yet the use of pressure cuffs in such cases is deemed impractical. Moreover, the regional PWV do not account for wave reflection and distance approximation of artery path length. To address this, researchers have introduced the concept of measuring local pulse wave velocity, crucial for obtaining a comprehensive understanding of blood vessels. local PWV measurement relies on simultaneous acquisition of the propagating blood pulse waves from two distinct locations at a known separation distance ($\Delta X$) within a small arterial segment. The time taken for the blood pulse wave to travel $\Delta X$ distance (pulse transit time [$\Delta T$]), is obtained as the time delay between an identifiable point (fiducial point) on the proximal and distal pulse cycle pair. Following the fundamental distance-time equation, local PWV is then calculated as Local PWV = $\Delta X/\Delta T$.

*1) Carotid Stiffness:* The algorithm used for automatic identification of the common carotid artery (CCA) using A-mode ultrasound in ARTSENS was evaluated. The algorithm was aimed at automating the identification of CCA walls to facilitate user-friendly operation without prior ultrasound knowledge to avoid the need of expert operators. The algorithm's speed of processing was found to be linearly dependent on the number of frames per measurement and independent of jump values. Simulation studies proved the algorithm's ability to reject blank frames and frames without the CCA, achieving a high HIT rate and low FALSE rate at low Signal to noise ratio (SNR) levels. Initial feasibility was tested in-vivo studies, the algorithm demonstrated a mean HIT rate of nearly 70% with a minimal FALSE rate when using more than 10 frames per measurement and higher jump values. The algorithm's performance, especially in simulation and in vivo studies, demonstrates its potential as a key component of the ARTSENS non-invasive screening tool for cardiovascular disease risk assessment [6].

A novel method and system for non-invasive, cuffless, and calibration-free evaluation of local carotid pulse pressure (ΔP) using a bi-modal arterial compliance probe was investigated. It consisted of two identical magnetic plethysmograph (MPG) sensors. The study includes in vitro verification using an arterial flow phantom and in vivo validation on 22 normotensive human subjects. The proposed technique uses a mathematical model for real-time evaluation of local ΔP based on simultaneously measured local pulse wave velocity (PWV) and arterial dimensions. The results showed a strong correlation between estimated and reference ΔP, with no significant bias. The results support the reliability and accuracy of the prototype device in capturing variations in carotid local ΔP, local PWV, and arterial dimensions induced by physical innervations [7].

*2) Magnetic Plethysmographic Transducers:* A single magnetic plethysmograph (MPG) transducer for the measurement of local pulse wave velocity (PWV) was assessed. The design aims to be small, reliable, and capable of capturing a large number of cardiac cycles. The proposed MPG transducer consists of a Hall effect sensor and a permanent magnet, with a strap-based design for reliable capture of blood pulse waveforms. The signals from the sensors are processed with an algorithm for local PWV measurement. The algorithm involves identifying critical points on the pulse waveforms and calculating the time delay between them to estimate the local PWV. The experimental validation involved in-vivo measurements on volunteers, which demonstrated the transducer's ability to capture carotid pulse waveforms continuously. Local PWV was measured on 15 volunteers, and the results were found to be consistent and repeatable. Additionally, this study's result also shows in accordance to previous works showing positive correlation between local PWV and diastolic pressure, systolic pressure, and mean arterial pressure. The results indicate that the proposed transducers could provide reasonably accurate measurements of local PWV and enable continuous BP monitoring without the need for a cuff [8].

Another study involved the development of a novel dual-element magnetic plethysmograph (MPG) arterial compliance probe, along with hardware implementation and measurement algorithms for local PWV. The pulse wave analysis algorithm used in the study involved identifying fiducial points within the pulse cycle and evaluating local PWV from the systolic rising point and neighbourhood of the systolic peak. An in-vivo validation study was conducted on 15 healthy volunteers, demonstrating the feasibility of local PWV evaluation from multiple fiducial points. The study found that local PWV obtained from the neighbourhood of the systolic peak was consistently higher than that obtained from the systolic rising point, demonstrating the variation in local PWV over the cardiac cycle. The coefficient of variation of the ratio of local PWV values at the two distinct fiducial points was less than 7%, indicating measurement reliability. The results demonstrate the potential for cuff-less, calibration-free evaluation of BP parameters from superficial arteries in a beat-by-beat manner [9].

*3) Photoplethysmographic Transducer:* The development and validation of a novel single-source Photoplethysmograph (PPG) transducer for the measurement of local pulse wave velocity (PWV) and detection of arterial blood artery in the carotid artery was evaluated. The transducer is designed to acquire dual blood pulse waveforms from the carotid artery continuously, enabling accurate local PWV measurement. The validation process consists of testing the transducer on an arterial flow phantom and conducting an in-vivo study involving 17 healthy volunteers. The results demonstrate the transducer's successful acquisition of blood pulse waveforms and its ability to measure carotid local PWV with notable repeatability and reliability. Importantly, the study establishes a positive correlation between carotid local PWV and brachial blood pressure parameters, indicating the potential use of local PWV as a surrogate parameter for cuffless blood pressure measurement systems. Overall this study highlights its clinical feasibility and potential use in calibration-free cuffless blood pressure monitoring devices using single source PPG transducer [10].

Current methods for local PWV involve costly technologies such as optical sensors, Doppler ultrasound, and MRI, requiring specialized expertise. However, transducer-based technology emerges as a promising, low-cost alternative for local PWV measurement, offering potential solutions to enhance accessibility and practicality in assessing vascular health.

*C. Prosthetic devices*

The design and implementation of an implantable microstimulator chip designed for stimulating electrodes in a retinal prosthetic device was evaluated. The microstimulator chip is capable of delivering biphasic current pulses of up to 600μA amplitude to a maximum load of 10kΩ. It operates at a low voltage headroom of 0.5V and uses an active feedback setup to ensure minimal gain variations for output currents. The active feedback current mirrors are used to minimize voltage headroom, and the demultiplexing units are integrated within the output stages of the driver for efficiency. Variable amplification schemes are employed to ensure high resolution across the entire stimulus range, and charge cancellation circuits are implemented to prevent charge buildup and ensure safety of the tissue. The chip's performance was tested for cathodic and anodic currents, with the results showing a matching of within 1.3%. The prototype chip, fabricated in AMI 1.6μm bulk CMOS technology, demonstrated linear output characteristics and efficient stimulus generation over a wide range. The design is also capable of producing biphasic stimulus waveforms and effectively discharging the electrodes to prevent charge build-up. Thus, the chip is designed to stimulate electrodes in order to create visual sensation in patients with retinitis pigmentosa and age-related macular degeneration [11].

The development of a test chip that will be used to evaluate a hermetic and biocompatible package for the driving CMOS circuitry of a retinal prosthesis was assessed. The methodology outlines the chip's design and functionality, including its suitability for hermeticity testing and the inclusion of moisture sensors to assess the long-term effect of moisture on electronic devices. The test procedure involves accelerated soak testing at higher temperatures to determine failure modes and mean time-to-failure under different encapsulation techniques. The results indicate that parylene-coated metal lines showed minimal impedance change during heated isotonic saline soak tests, suggesting perylene's effectiveness as a water and salt barrier for the CMOS circuitry. The combination of metal and parylene is estimated to maintain package integrity in vivo for over 10 years. Overall, the test chip's design, the testing procedure, results of soak tests, and aims to develop a robust hermetic sealing technique for biomedical implants. The use of parylene and metal for encapsulation is being explored to prevent failure modes in CMOS circuitry [12].

*D. Wearable devices*

Neuroscientists employ neural recording systems to monitor the behavior of live animals such as birds, reptiles, rats, and primates. In those experiments, animals under testing are usually anesthetized or bound by tethered wires, which imposes a great limitation on the behavior being observed and the information quality of the signals. Hence small wireless system which can deliver signal without compromising on the quality is required.

A 4-channel wearable wireless neural recording system design was assessed for monitoring live animal biopotentials. The head stage records from 4 channels and is powered by two button cell batteries, enabling continuous operation for at least 15 hours. The receiver board includes an FM demodulator and an amplifier, with the data captured by a National Instruments data acquisition device and transmitted to a PC via USB. The chip's signal path is fully differential, and it offers adjustable channel gain and bandwidth, enabling precise recording of neural signals. The system's testing involved successful ex-vivo and in-vivo recordings from a dissected snail brain and a live rat, respectively. The performance was compared with previous works, highlighting its advantages in terms of weight, dimension, power consumption, amplifier technology, and data telemetry. The system's successful testing on both dissected and live animal subjects validates its effectiveness for research and monitoring purposes of small animals [13].

An algorithm for real-time heart rate monitoring using wrist-type photoplethysmographic (PPG) signals was evaluated for humans. The algorithm uses Multiple Reference ADaptive noise cancellation technique– termed here as 'MURAD'. It uses multiple reference signals, such as 3-axis accelerometer data and the difference signal between two PPG signals, to clean the PPG signal and estimate heart rate. The algorithm also includes a peak verification stage to ensure accurate heart rate estimation. Experimental results demonstrate the algorithm's effectiveness in reducing motion artifacts and accurately estimating heart rate, even during extreme physical activities such as weightlifting, swimming, and boxing. The algorithm outperforms other state-of-the-art methods in terms of average absolute error and absolute error percentage for both training and testing datasets. Overall, the proposed algorithm provides a promising solution for robust heart rate monitoring using wrist-type PPG signals, especially during severe motion artifact conditions. It has the potential to be applied in various wearable devices and mobile applications for real-time heart rate monitoring in challenging environments [14].

III. DISCUSSION

All stiffness devices demonstrate validity, reproducibility, and repeatability by comparing to different reference devices, establishing a universal reference system for non-invasive devices remains challenging for clinical application. Brachial PWV measurement, for instance, validated against aortic PWV, acknowledges the difficulty in accounting for differences in path length and artery types present in these lengths. Aortic and brachial PWV are not reproducible when measured using applanation tonometry. Participants are in a supine position during measurements, except for ARTSENS, which has demonstrated reproducibility even in a sitting position, making it an ideal candidate for screening in resource-controlled settings. The time required needs to be mentioned for a full measurement, as observed with ARTSENS, is vital for considering a device as a screening tool. However, these devices have not accounted for special populations, limiting their use in yet-to-be-studied groups. Overall, all these devices confirm the validity and reproducibility of non-invasive PWV measurements, indicating their potential clinical significance in assessing vascular damage and cardiovascular risk across various settings.

The evaluated implantable microstimulator chip demonstrates promising advancements for retinal prosthetic devices, offering high-resolution visual stimulation and safety features. The accompanying test chip assesses a hermetic sealing technique using parylene-coated metal lines, showing potential for long-term reliability in CMOS circuitry for retinal prostheses. These innovations represent significant progress in developing effective and durable solutions for patients with retinitis pigmentosa and age-related macular degeneration.

IV. CONCLUSION

The findings suggest a growing trend in the development and utilization of non-invasive technologies for monitoring and diagnosing health conditions related to the cardiovascular and ophthalmic systems. In the cardiovascular domain, researchers have made progress in creating devices that provide accurate and real-time data without the need for invasive procedures, contributing significantly to early detection, continuous monitoring, and effective management of cardiovascular diseases. The integration of innovative technologies in biomedical research has led to developments in retinal prosthetics and hermetic packaging for implantable devices. The successful design and implementation of an implantable microstimulator chip tailored for retinal prosthetic applications offer a promising solution for patients with retinitis pigmentosa and age-related macular degeneration. The comprehensive hermetic packaging integrates analog and digital components along with environmental sensors, suggesting the potential for a robust hermetic sealing technique that could maintain package integrity in vivo for extended periods, exceeding 10 years. Overall, the synthesis of findings from these research papers emphasizes the transformative potential of non-invasive biomedical devices in the realms of cardiovascular and ophthalmic healthcare. As technology continues to evolve, it is evident that these non-invasive solutions have the potential to reshape diagnostic and monitoring practices, ultimately leading to more effective and patient-friendly healthcare interventions in these critical medical domains.